# Multimodal active speaker detection and virtual cinematography for video conferencing

*Ross Cutler[1], Ramin Mehran[2*], Sam Johnson[3*], Cha Zhang[1], Adam Kirk[4*], Oliver Whyte[4*], Adarsh Kowdle[5*]*

[1]Microsoft Corp, Redmond, WA [2]Zillow, Seattle, WA [3]Facebook, Menlo Park, CA
[4]Omnivor, Seattle, WA [5]perceptiveIO, San Francisco, CA
{rcutler,chazhang}@microsoft.com, {rmehran,sammoj,adam.g.kirk,oawhyte,adarsh.kp}@gmail.com

## ABSTRACT

Active speaker detection (ASD) and virtual cinematography (VC) can significantly improve the remote user experience of a video conference by automatically panning, tilting and zooming of a video conferencing camera: users subjectively rate an expert video cinematographer's video significantly higher than unedited video. We describe a new automated ASD and VC that performs within 0.3 MOS of an expert cinematographer based on subjective ratings with a 1-5 scale. This system uses a 4K wide-FOV camera, a depth camera, and a microphone array; it extracts features from each modality and trains an ASD using an AdaBoost machine learning system that is very efficient and runs in real-time. A VC is similarly trained using machine learning to optimize the subjective quality of the overall experience. To avoid distracting the room participants and reduce switching latency the system has no moving parts – the VC works by cropping and zooming the 4K wide-FOV video stream. The system was tuned and evaluated using extensive crowdsourcing techniques and evaluated on a dataset with N=100 meetings, each 2-5 minutes in length.

## 1. INTRODUCTION

Video conferencing is widely used for remote collaboration, and many conference rooms in businesses have a video conferencing system installed to help facilitate remote collaboration for employees. A few commercial video conferencing systems (e.g., [1], [2]) have active speaker detection (ASD) to track the active speaker and give an enhanced video experience to the far-end (e.g., [3], [4], [5]) but the vast majority of video conferencing systems do not. ASD allows the far-end participants to see who is currently speaking, which is especially useful when the conference room is large or the remote video is rendered on a small display due to small screen size, small render size, or limited bandwidth. For example, in Figure 1 a 480p video stream is viewed remotely on a smartphone, captured from a 2160p video conferencing camera. If the original video is scaled to display at 480p or even in the 2160p video is shown on a smartphone, the faces are too small to recognize reliably, greatly diminishing the value of video for remote collaboration. In contrast, the cropped video stream is much more informative.

Some of the limitations of existing ASD solutions are (1) available commercial systems have high latency (e.g., >2s), (2) the systems use multiple mechanical pan-tilt-zoom (PTZ) cameras and/or large 2D microphone arrays, which can be distracting, (3) the systems are expensive, limiting the number of deployments. In this paper, we describe a new video conferencing system with ASD that addresses these issues. In particular, (1) our system achieves <200ms ASD+VC latency (2) the system contains no moving parts or large 2D microphone arrays; it leverages a depth camera to reduce physical size and noise, thus avoiding distractions within the conference room, and (3) the system uses a single 4K wide-FOV camera to replace multiple expensive PTZ cameras, which significantly reduced cost, and (4) our system achieves subjective performance within 0.3 MOS (using a 1-5 scale) of an expert virtual cinematographer.

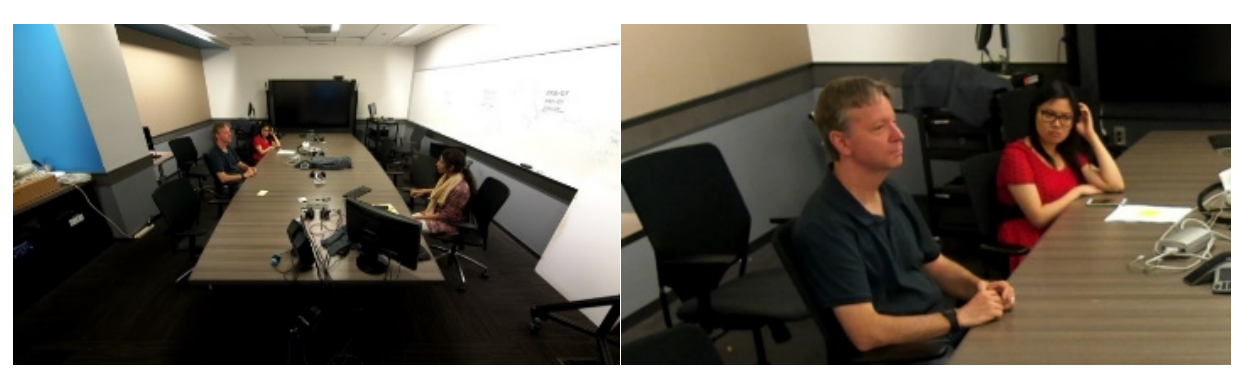

**Figure 1: Left: Original 2160p video; Right: Cropped 480p image from a video conference**

## 2. RELATED WORK

The ASD and VC developed by Zhang et. al [5] is the most related work to ours. In that work, a one-dimensional ASD and VC were trained using AdaBoost. However, that system only had to estimate azimuth, not zoom or elevation, which greatly simplified the problem. Commercially available systems such as [1] and [2] solve the PTZ problem using large two-dimensional microphone arrays to estimate elevation and depth. However, these systems have significant delays in changing to the active speaker (>2s). Our system requirements are to switch to the active speaker < 200ms with no secondary camera nor large 2D microphone array. Systems with digital PTZ such as [1] and [6] also have significant delays (>2s) and do not handle large rooms.

An early work in multimodal ASD used TDNN [7], and there has been significant recent work using DNN-based ASD (e.g., [8], [9], [10], [11], [12], [13], [14]). There is now a large dataset created for this task [15] with an ASD competition [10]. However, we are not aware of any ASD (DNN or otherwise) that does low-latency accurate PTZ without large 2D SLL arrays. In addition, this is the first study we are aware of that provides a subjective evaluation of the VC performance compared to a human expert.

---

[*] Work done while at Microsoft Corp.

## 3. SYSTEM OVERVIEW

Our system uses the following multimodal sensors:

- A 4-element linear microphone array logarithmically spaced as shown in Figure 4 with a total width of 215mm.
- A depth camera with a 512x424 resolution and 0.5-10m working range.
- A 4K (3840x2160) RGB video camera with a 100° HFOV.

The microphone array uses unidirectional microphones and is sampled at 16 kHz. The depth camera and RGB camera are synchronized with the same start-of-frame signal. Video and audio frames use a common timestamp to facilitate synchronization between audio/video modalities.

The system dataflow is shown in Figure 2. The microphone array is processed with the sound source localization (SSL) method described in [17]. SSL features are estimated from the SSL probability distribution function (PDF). A set of 15 such features are defined using local $L^l$ and global $L^g$ minima and maxima over the PDF, described in Table 1. The depth camera is used to estimate the location of the conference room table to limit the range for the PTZ locations. In addition, depth features based on Haar-like wavelets (Figure 5) and short and long-range motion features as estimated in [5] are estimated from the depth camera. Finally, the depth camera is used to estimate the zoom used in the VC. The RGB video camera is used to estimate similar motion and video features; in additional a face detector is used to estimate face rectangles. All the above features are fed into an AdaBoost-based ASD.

The system architecture is shown in Figure 3. The system runs on a single-core of a 2 GHz Intel i5 CPU based embedded PC. There are 4 threads in the ASD and VC that process the features and evaluate the trained AdaBoost ASD and state-machine VC.

The VC is implemented as a state-machine described in Figure 6. The VC has four states: (1) stationary, (2) update target for a global view (zoom out), (3) update the target for a cut (pan/tilt/zoom), and (4) update the window. The parameters have been tuned to maximize user ratings as described in later sections.

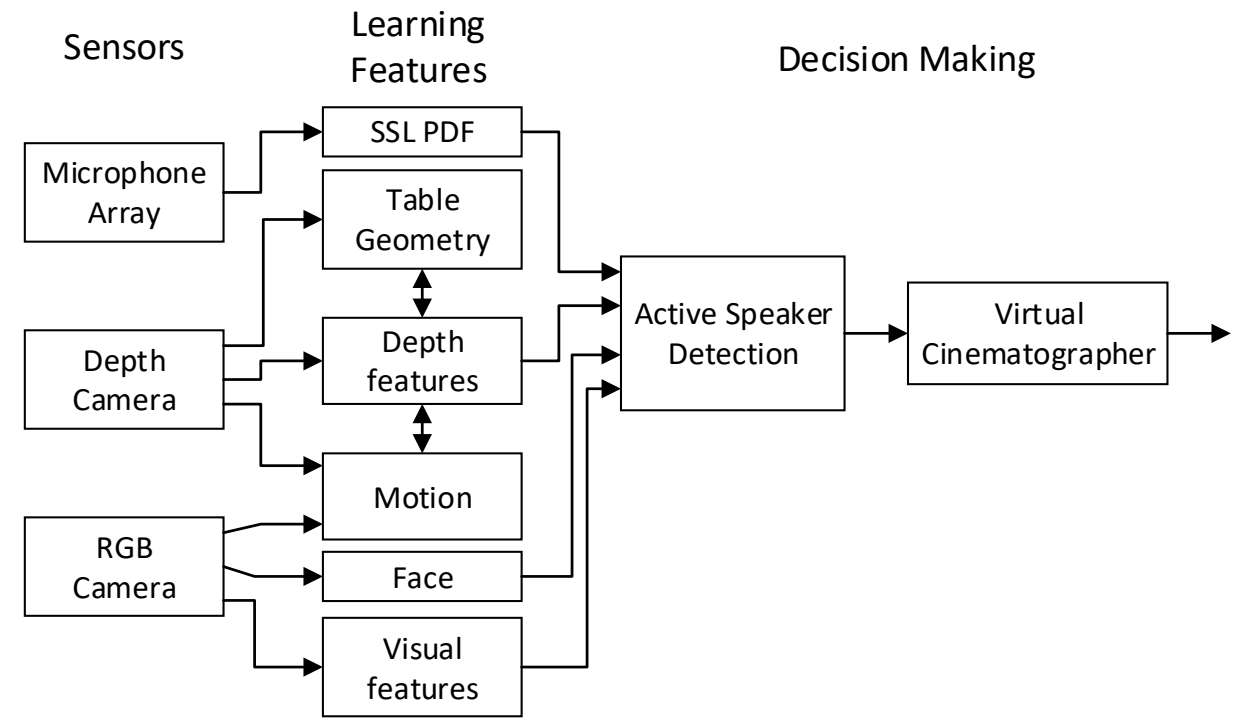

**Figure 2: System dataflow**

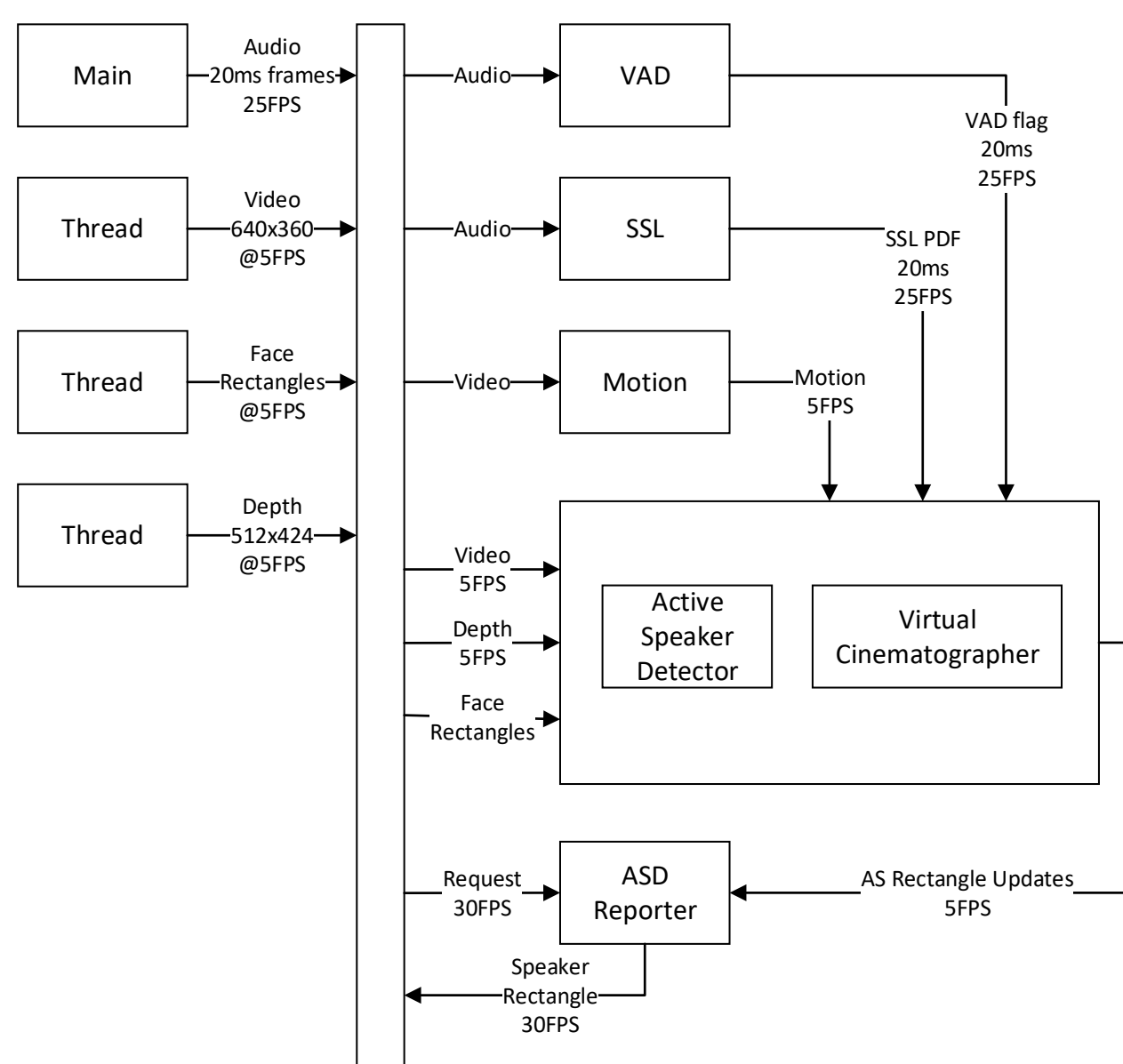

**Figure 3: System Architecture**

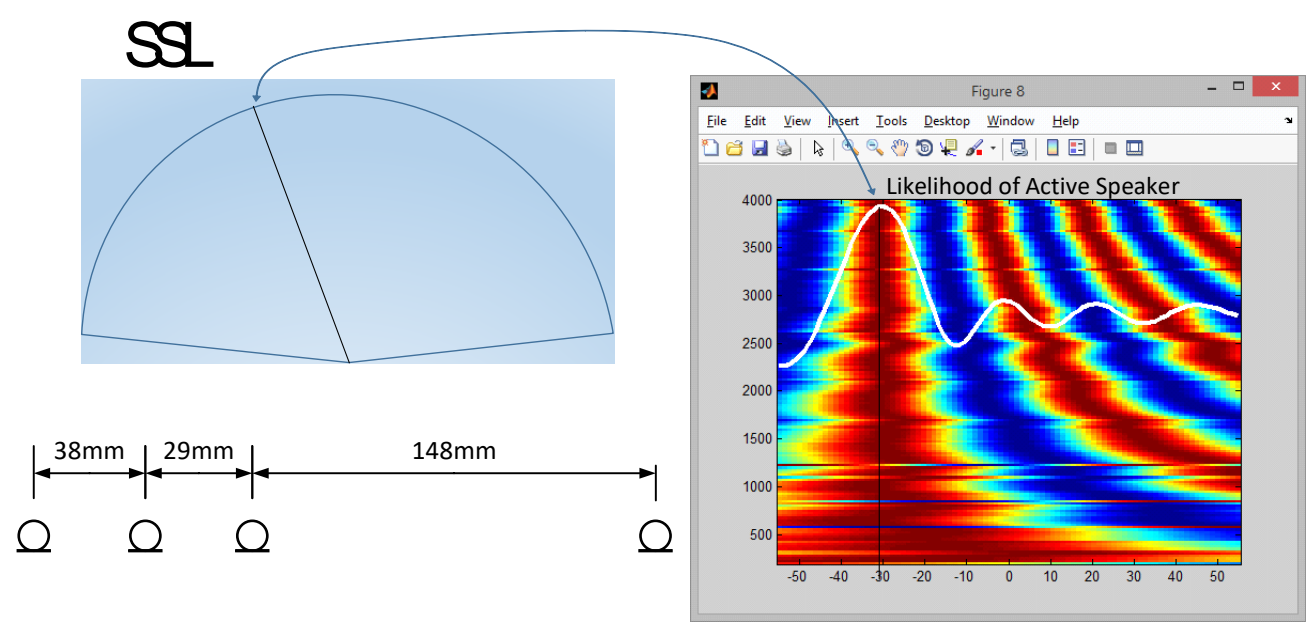

**Figure 4: Sound Source Localization PDF**

| 1. $\frac{L^l_{max}-L^g_{min}}{L^g_{max}-L^g_{min}}$ $\frac{L^l_{max}-L^g_{min}}{L^g_{max}-L^g_{min}}$ | 2. $\frac{L^l_{min}-L^g_{min}}{L^g_{max}-L^g_{min}}$ | 3. $\frac{L^l_{avg}-L^g_{min}}{L^g_{max}-L^g_{min}}$ |
|---|---|---|
| 4. $\frac{L^l_{mid}-L^g_{min}}{L^g_{max}-L^g_{min}}$ | 5. $\frac{L^l_{max}}{L^l_{min}}$ | 6. $\frac{L^l_{max}}{L^l_{avg}}$ |
| 7. $\frac{L^l_{min}}{L^l_{avg}}$ | 8. $\frac{L^l_{min}}{L^l_{avg}}$ | 9. $\frac{L^l_{max}-L^l_{min}}{L^l_{avg}}$ |
| 10. $\frac{L^l_{max}}{L^g_{max}}$ | 11. $\frac{L^l_{min}}{L^g_{max}}$ | 12. $\frac{L^l_{avg}}{L^g_{max}}$ |
| 13. $\frac{L^l_{mid}}{L^g_{max}}$ | 14. $\frac{L^l_{max}-L^l_{min}}{L^g_{max}}$ | 15. $L^g_{max} - L^l_{max} < \epsilon$ |

**Table 1: Audio SSL features**

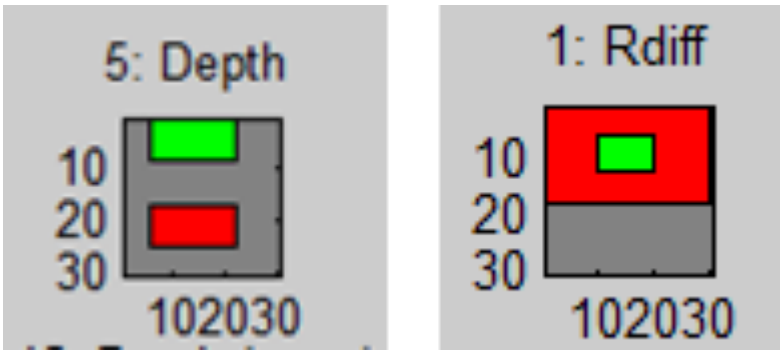

**Figure 5: Example video features, Haar-like wavelets**

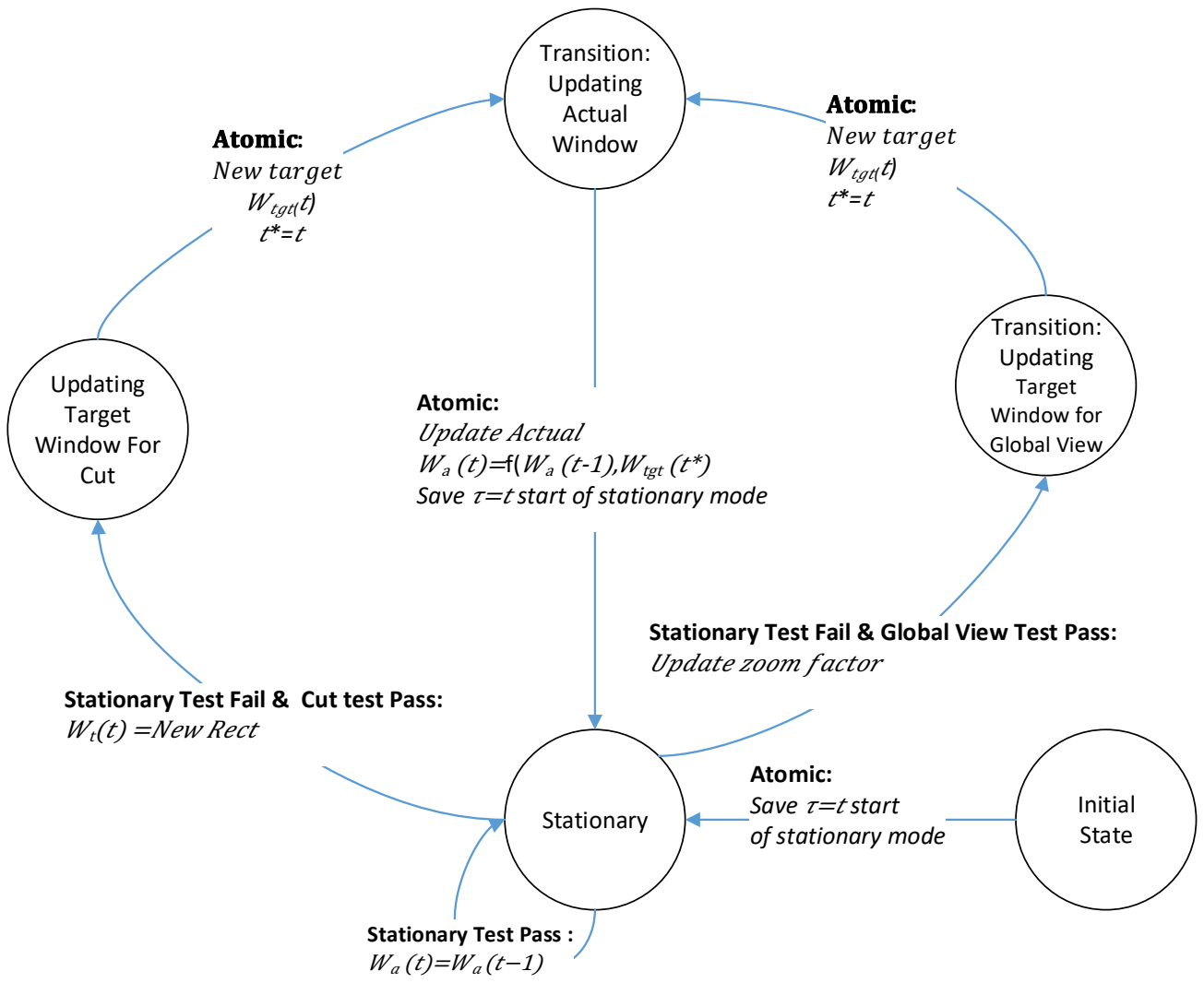

**Figure 6: Virtual Cinematographer state machine**

## 4. DATA COLLECTION

The system we have designed requires a significant amount of training and test data for supervised learning of the ASD and VC. The primary goal for the data capture of meetings is to capture a large variety of meeting data, similar to that which the device will see in actual usage.

We estimated based on previous work [5] that we would need at least 100 meetings of 5 minutes each for the training and testing of the ASD and VC (we later used cross-validation to check it was sufficient). The requirements for the collection are:

- Coverage of the space of different:
  - Rooms and meeting types
  - Distribution and number of people
  - Speaker variations and distractions
  - Lighting and appearance variations
- Video frames must be labeled with:
  - Bounding boxes around all heads in view
  - Current active speaker

An example set of meetings from this dataset are shown in Figure 7.

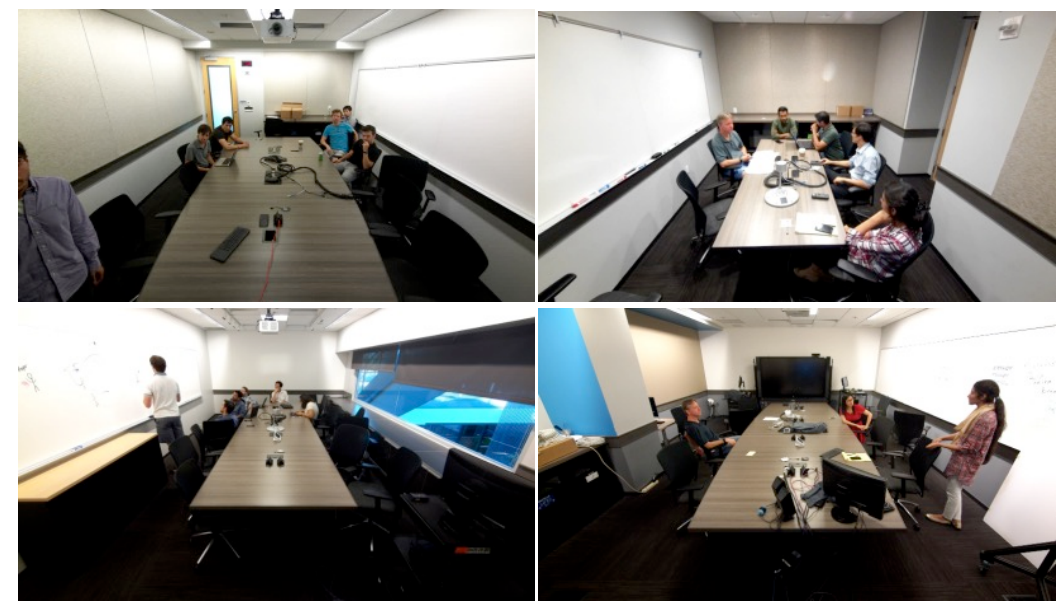

**Figure 7: Example meetings in the dataset**

We labeled the dataset at 5 FPS, giving 150K total frames to label. Assuming 15 seconds to label the 4-20 heads in each frame this would be **26 straight days of labeling.** To speed this up we used crowdsourcing for labeling. The key challenge in this solution is getting high-quality output from unskilled workers at low pay ($0.03/frame). We take the following process to maximize worker quality: (1) offline training with feedback, (2) qualification of raters, (3) online training with feedback, (4) online spam checking. Finally, we take multiple annotations per frame, repeating until 2 users agree for each box. Many frames of video are essentially the same – most meeting participants move very little. We use a state-of-the-art tracker to interpolate the location of participants between labeled frames (Figure 8). Working on an active learning approach we: (1) start with a very sparse label set, (2) track forward and backward, (3) when the tracker posterior drops below a set threshold for any frame, we request labeling. For the 150K frames of collected data, we also wish to know who is actively speaking at that moment. We follow the same approach as for bounding boxes. Here we show annotators a 4-second video with the middle frame being the frame of interest. Annotators are asked to pick the speaker at the moment bounding boxes are flashed.

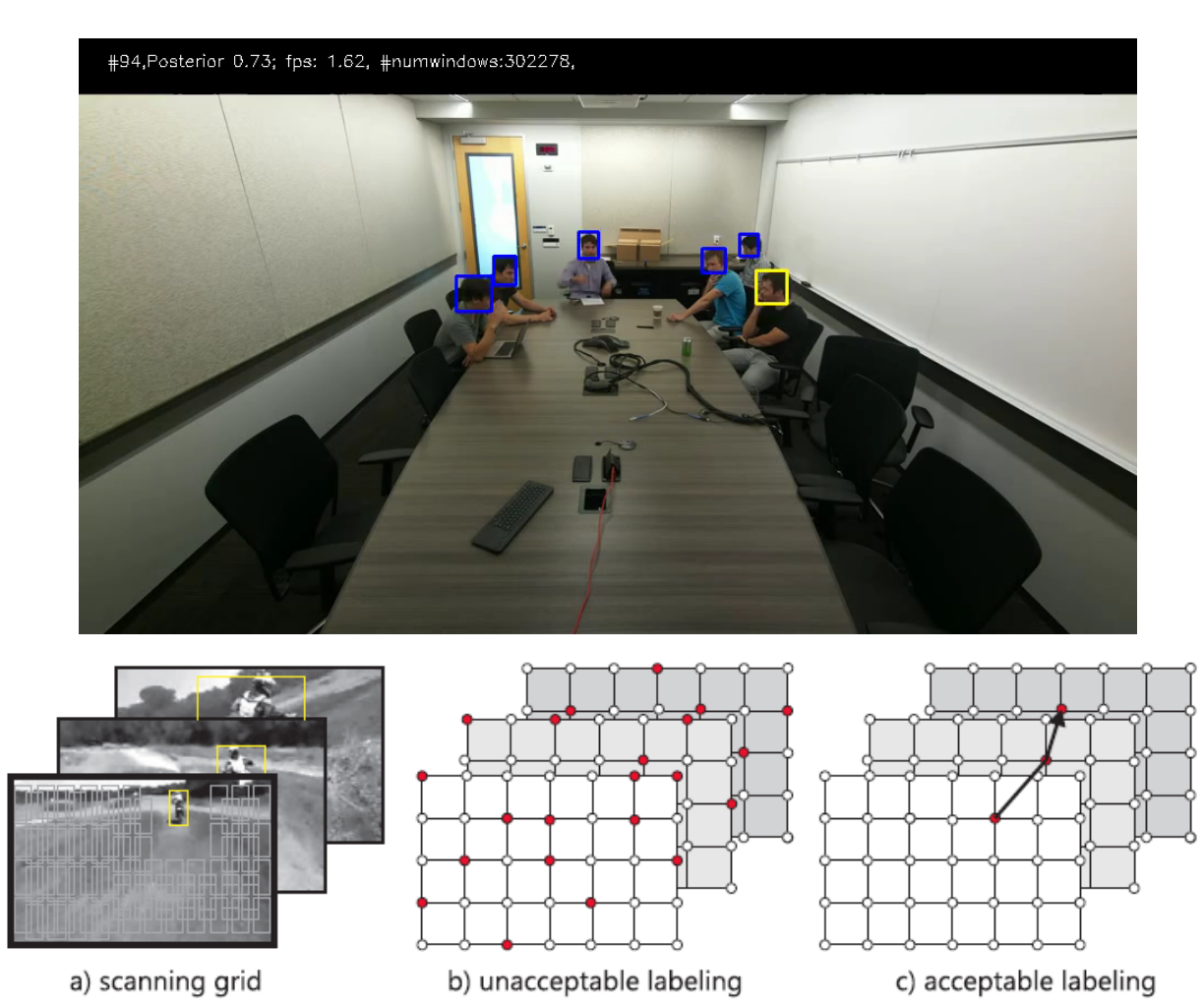

**Figure 8: The tracker used in crowdsourced labeling of data**

## 5. TRAINING AND TESTING

The system was trained and tested using cross-validation. To determine whether various components needed further improvements we substituted those components with ground truth data to check if there were any overall improvements. For example, to determine if the SSL algorithm needed improvement we substituted the estimated SSL with the ground truth SSL and measured the system performance difference (there was little difference). The same was done for the ASD and face detector.

The statistics of the ASD features used are given in Figure 9. This shows that the depth (and the "normalized" depth which imputes missing values) dominates the number of AdaBoost features. Rdiff is the long-term image differences, and Diff is the short-term image differences.

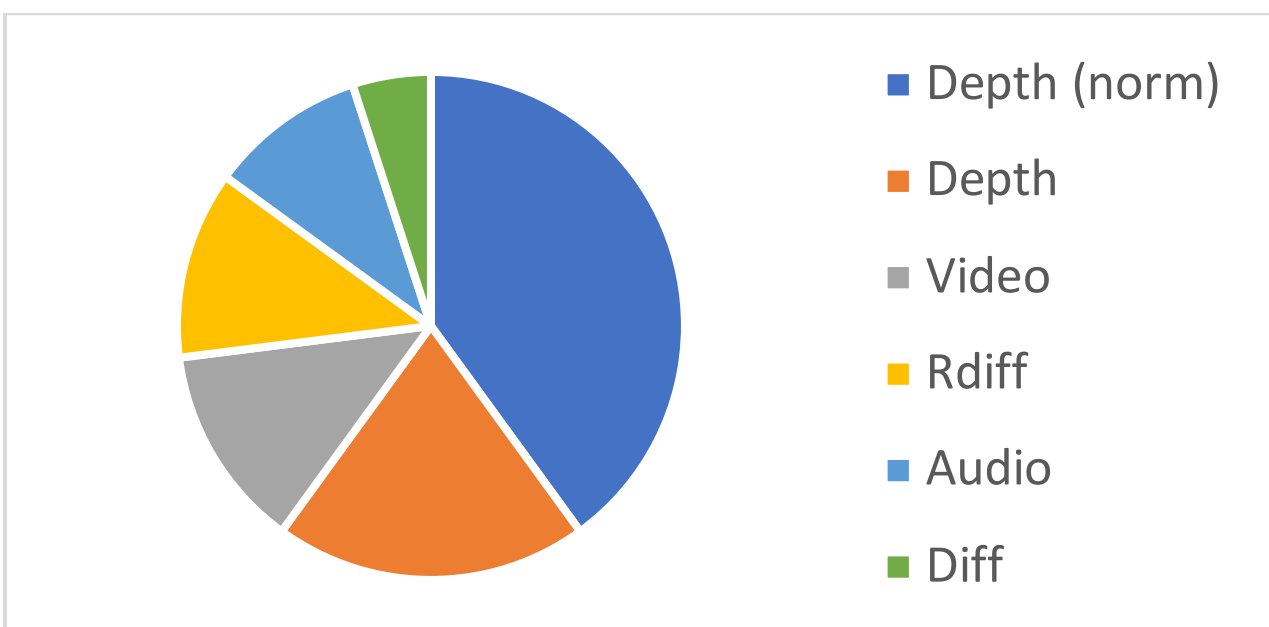

**Figure 9: Statistics of used features in AdaBoost ASD**

## 6. RESULTS AND DISCUSSION

To evaluate the system we utilize both objective and subjective metrics. For the objective metrics, we defined four key performance indicators (KPIs) that characterize the performance of the ASD and VC. These KPIs are defined below:

- ASD Speaker Detection Rate (SDR): Out of frames containing speakers, how often the ASD found a speaker
- ASD Person Detection Rate (PDR): Out of frames containing speakers, how often the ASD found a person
- ASD False Negative Rate (FNR): Out of all frames containing speakers, how often the ASD didn't fire while someone was speaking
- VC Acceptable Speaker Rate (ASR): How often the VC produced an "acceptable" crop. The crop should show a speaker, or show a person if no one is speaking

The results are shown in Table 2. Small rooms fit 6 or fewer people, medium rooms fit 7-16 people, and large rooms are >16 people. Overall the ASD detects the correct speaker 98.3% of the time, and when someone is speaking a person is selected 99.2% of the time (a non-person is selected just 0.8% of the time). The VC performance is better for small rooms (VC ASR = 96.5%) than large rooms (VC ASR = 90.4%), which is expected.

| Room | ASD SDR | ASD PDR | ASD FNR | VC ASR |
|---|---|---|---|---|
| Small | 97.1% | 98.5% | 0.1% | 96.5% |
| Medium | 99.4% | 99.5% | 2.5% | 91.5% |
| Large | 97.6% | 99.0% | 0.5% | 90.4% |
| **Total** | **98.3%** | **99.2%** | **1.3%** | **91.4%** |

**Table 2: ASD and VC results**

To determine what KPI criteria is needed to be subjectively good enough we performed a subjective test using crowdsourcing. A selection of N=100 one minute videos was edited by an expert cinematographer and these videos were compared with the VC edited videos using the survey form shown in Figure 10. The results are shown in Table 3 and show that the VC gets within 0.3 Mean Opinion Score (MOS) [18] of an expert cinematographer, which was sufficient for our requirements.

Most of the complaints in the VC subjective test had to do with the mishandling of meetings when the whiteboard was used, in which the VC did not show what the remote person was writing, unlike the expert cinematographer.

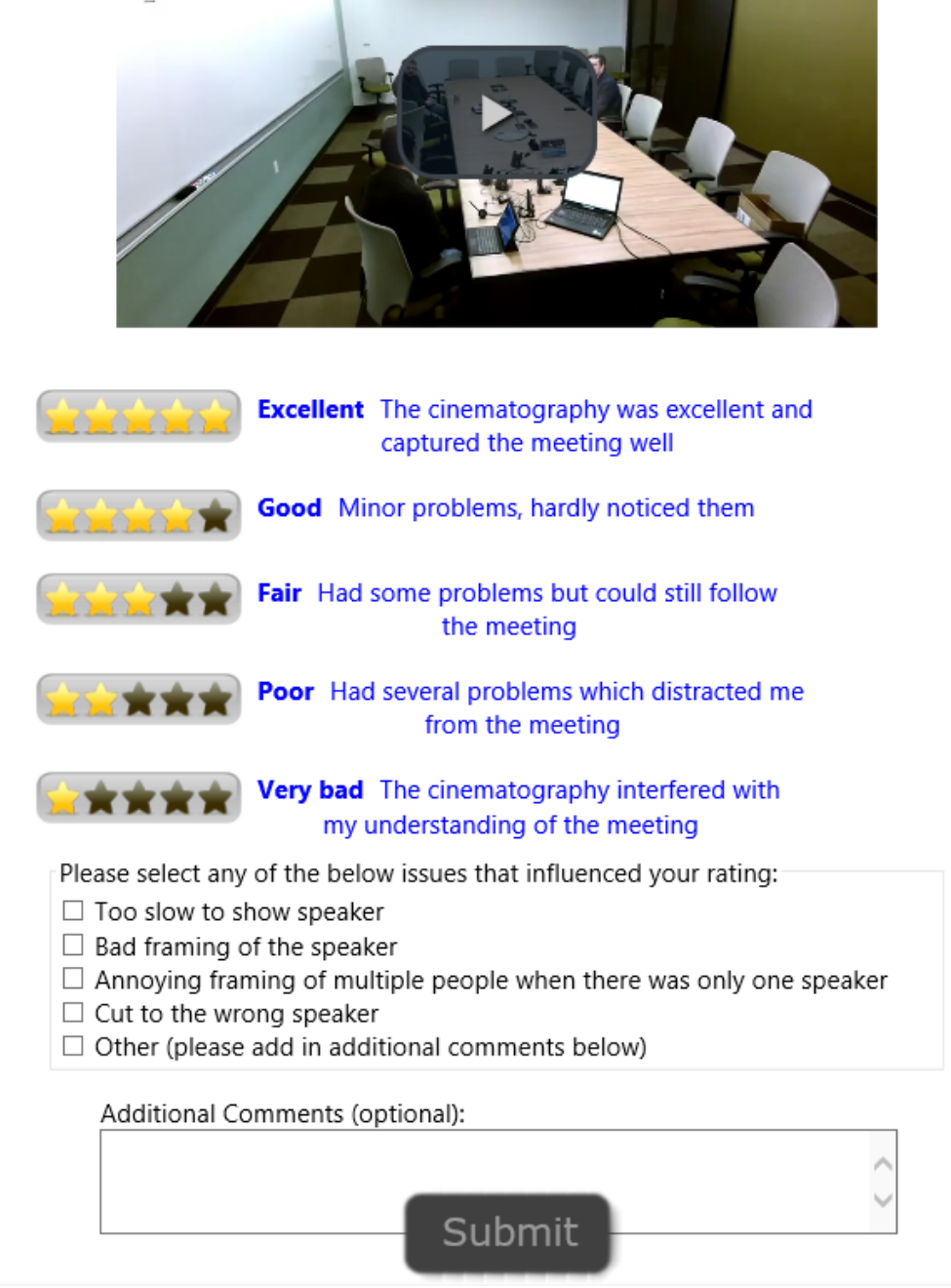
Please watch the entire video clip below, and rate the quality of the cinematography, i.e. how well can you follow the meeting given what is displayed within the video frame. **If your video appears green please refresh the page.**

**Excellent** The cinematography was excellent and captured the meeting well

**Good** Minor problems, hardly noticed them

**Fair** Had some problems but could still follow the meeting

**Poor** Had several problems which distracted me from the meeting

**Very bad** The cinematography interfered with my understanding of the meeting

Please select any of the below issues that influenced your rating:
- ☐ Too slow to show speaker
- ☐ Bad framing of the speaker
- ☐ Annoying framing of multiple people when there was only one speaker
- ☐ Cut to the wrong speaker
- ☐ Other (please add in additional comments below)

Additional Comments (optional):

Submit

**Figure 10: Survey form used to rate the VC for subjective evaluation**

| | MOS |
|---|---|
| Expert Cinematographer | 4.1 ± 0.1 |
| VC | 3.8 ± 0.1 |

**Table 3: VC subjective results with 95% confidence interval**

## 7. CONCLUSIONS

We have described an end-to-end system for an ASD and VC that uses multisensory input to perform within 0.3 MOS of a human expert cinematographer. This is done with low latency and in a compact form factor with no moving parts to distract the near-end participants. Future enhancements can be made to improve the performance including:

- Improving the voice activity detector to reduce incorrect jumps to, for example, squeaky chairs, paper shuffling, etc.
- Improving whiteboard handling to include the last say one minute of what was written on the whiteboard in addition to the person writing on the whiteboard
- Using deep learning for better audio/video features into AdaBoost, and perhaps using a DNN instead of AdaBoost for improved performance.